\documentclass[showpacs,amsmath,
twocolumn,
aps,prb]{revtex4}
\usepackage{bm} 
\usepackage{graphicx}
\usepackage{dcolumn}
\usepackage[colorlinks=true, pdfstartview=FitV, linkcolor=blue, 
            citecolor=blue, urlcolor=blue]{hyperref}

\begin{document}

\title{Force-detected nuclear magnetic resonance: Recent advances and
  future challenges}

\author{M.  Poggio$^{1}$ and C. L. Degen$^{2}$}
\affiliation{$^1$Department of Physics, University of Basel, Klingelbergstrasse 82, 4056 Basel, Switzerland \\
  $^2$Department of Chemistry, Massachusetts Institute of Technology, 77 Massachusetts Avenue, Cambridge, MA 02139, USA}
\date{\today}

\begin{abstract}
  We review recent efforts to detect small numbers of nuclear spins
  using magnetic resonance force microscopy. Magnetic resonance force
  microscopy (MRFM) is a scanning probe technique that relies on the
  mechanical measurement of the weak magnetic force between a
  microscopic magnet and the magnetic moments in a sample.  Spurred by
  the recent progress in fabricating ultrasensitive force detectors,
  MRFM has rapidly improved its capability over the last decade.
  Today it boasts a spin sensitivity that surpasses conventional,
  inductive nuclear magnetic resonance detectors by about eight orders
  of magnitude.  In this review we touch on the origins of this
  technique and focus on its recent application to nanoscale nuclear
  spin ensembles, in particular on the imaging of nanoscale objects
  with a three-dimensional (3D) spatial resolution better than 10 nm.
  We consider the experimental advances driving this work and
  highlight the underlying physical principles and limitations of the
  method.  Finally, we discuss the challenges that must be met in
  order to advance the technique towards single nuclear spin
  sensitivity -- and perhaps -- to 3D microscopy of molecules with
  atomic resolution.
\end{abstract}


\maketitle

\section{Introduction}
\label{Introduction}

Advances in the fabrication and measurement of microelectromechanical
systems (MEMS) and their evolution into nanoelectromechanical systems
(NEMS) have allowed researchers to measure astoundingly small forces,
masses, and displacements
\cite{Schwab:2005,ClelandBook:2002,Ekinci:2005,Blencowe:2004}.
Ultrasensitive mechanical sensors have been used in a variety of
experiments from the measurement of Casimir forces predicted by cavity
quantum electrodynamics \cite{Chan:2001}, to the testing of quantum
theories of gravity at nanometer length scales \cite{Smullin:2005}, to
the measurement of single electron spins \cite{Rugar:2004}.  Recently,
micromechanical detectors have provided the clearest picture to date
of persistent currents in normal metallic rings
\cite{Bleszynski:2009}.  At this stage, the measurement of forces at
the attonewton level has been reported by several teams of researchers
\cite{Stowe:1997,Mamin:2001,Rugar:2004,Naik:2006,Teufel:2009}.  At the
same time, mechanical displacement sensitivities are approaching the
standard quantum limit, i.e.\ the fundamental limit to position
resolution set by back-action effects \cite{Teufel:2009}.  NEMS
devices operated at very low temperatures are themselves approaching
the quantum regime, with thermal vibration energies only 100 quanta
above the zero-point motion of the resonators
\cite{Knobel:2003,LaHaye:2004,Naik:2006,Rocheleau:2009}.  The
availability of devices with such exquisite force, mass, and
displacement sensitivities has not only allowed for the study of a
wide class of condensed matter physics problems, but it has also led
to new high resolution nano- and atomic-scale imaging techniques.
Magnetic resonance force microscopy, which is reviewed here, has made
important contributions to all of these emerging research areas.

Magnetic resonance force microscopy (MRFM) combines the physics of
magnetic resonance imaging (MRI) with the techniques of scanning probe
microscopy.  In MRFM, a nanomechanical cantilever is used to sense the
tiny magnetic force arising between the electron or nuclear spins in
the sample and a nearby magnetic particle.  The MRFM technique was
proposed in the early days of scanning probe microscopy as a method to
improve the resolution of MRI to molecular lengthscales
\cite{Sidles:1991,SidlesPRL:1992}.  The visionary goal of this
proposal was to eventually image molecules atom-by-atom, so as to
directly map the three-dimensional atomic structure of macromolecules
\cite{SidlesRSI:1992}.  Such a ``molecular structure microscope''
would have a dramatic impact on modern structural biology, and would
be an important tool for many future nanoscale technologies.  While
this ultimate goal has not been achieved to date, the technique has
undergone a remarkable development into one of the most sensitive spin
detection methods available to researchers today.  Among the important
experimental achievements are the detection of a single electronic
spin \cite{Rugar:2004} and the extension of the spatial
resolution of MRI from several micrometers to below ten nanometers
\cite{DegenTMV:2009}.

In this review we discuss the improvements made to the MRFM technique
over the last four years, and present an outlook of possible future
developments.  Section \ref{Background} introduces the basics of MRFM
and provides a brief historical overview covering earlier work until
about 2006 (for a broader discussion of this work the reader is
referred to several reviews, for example Refs.
\cite{Sidles:1995,Nestle:2001,Suter:2004,Kuehn:2008,Berman:2006,Hammel:2007,Barbic:2009}).
Section \ref{nMRFMResearch} primarily focuses on work done by the
authors and collaborators while in the MRFM group at the IBM Almaden
Research Center.  We discuss the recent experimental advances that
allowed the measurement sensitivity to reach below 100 nuclear spins.
We highlight some of the results enabled by this progress, in
particular the imaging of individual virus particles and organic
nanolayers, both with three-dimensional (3D) resolutions below 10 nm.
We also consider two physical phenomena that become important at these
small length-scales: the role of statistical fluctuations in spin
polarization and the appearance of fast spin relaxation by coupling to
mechanical modes.  In Section \ref{Developments} we discuss promising
future directions aimed at improving the sensitivity of nuclear MRFM:
the development of improved magnetic tips, of novel nanomechanical
sensors, and of sensitive displacement transducers.  We conclude with
a comparison to other nanoscale imaging and spin detection techniques
in Section \ref{Othertechniques}, and an outlook of future
applications in Section \ref{Outlook}.

\section{Background}
\label{Background}

\subsection{Principle}
\label{Principle}

MRI and its older brother, nuclear magnetic resonance (NMR)
spectroscopy, rely on measurements of the nuclear magnetic moments
present in a sample -- magnetic moments arising from atomic nuclei
with non-zero nuclear spin.  In conventional magnetic resonance
detection, the sample is placed in a strong static magnetic field in
order to produce a Zeeman splitting between the nuclear spin states.
The sample is then exposed to a radio-frequency (rf) magnetic field of
a precisely defined frequency.  If this frequency matches the Zeeman
splitting (which at a given static field is different for every
non-zero nuclear spin isotope), then the system absorbs energy from
the rf radiation resulting in transitions between the nuclear spin
states.  From a classical point of view, the total nuclear magnetic
moment of the sample starts changing its orientation.  Once the rf
field is turned off, any component of the total moment remaining
perpendicular to the static field is left to precess about this field.
The precession of this ensemble of nuclear magnetic moments produces a
time-varying magnetic signal that can be detected with a pick-up coil.
The electric current induced in the coil is then amplified and
converted into a signal that is proportional to the number of nuclear
moments (or spins) in the sample.  In MRI this signal can be
reconstructed into a 3D image of the sample using spatially varying
magnetic fields and Fourier transform techniques.  The magnetic fields
produced by nuclear moments are, however, extremely small: more than
one trillion ($10^{12}$) nuclear spins are typically needed to
generate a detectable signal.

\begin{figure}\includegraphics[width=3.2in]{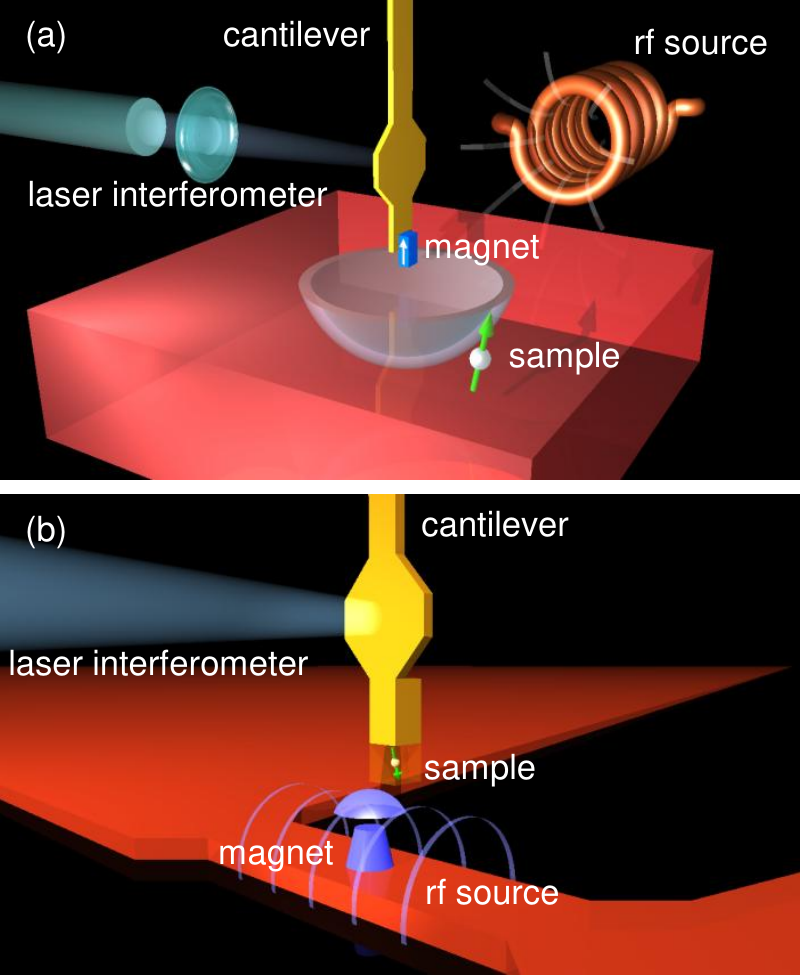}
  \caption{\footnotesize Schematics of an MRFM apparatus.  (a)
    corresponds to the ``tip-on-cantilever'' arrangement, such as used
    in the single electron MRFM experiment of 2004 \cite{Rugar:2004}.
    (b) corresponds to the ``sample-on-cantilever'' arrangement, like
    the one used for the nanoscale virus imaging experiment in 2009
    \cite{DegenTMV:2009}. In both cases the hemispherical region
    around the magnetic tip is the region where the spin resonance
    condition is met -- the so-called ``resonant slice''.}
\label{fig1}
\end{figure}

The MRFM technique attempts to improve on the poor detection
sensitivity of inductive pick-up coils by mechanically detecting the
magnetic forces produced by nuclear moments.  To grasp the basic idea
behind the method, imagine taking two refrigerator magnets and holding
them close together; depending on the magnets' orientation, they exert
either an attractive or repulsive force. In MRFM, a compliant
cantilever is used to sense the same magnetic forces arising between
the nuclear spins in a sample and a nearby nano-magnet. First, either
the sample (containing nuclear moments) or the nano-magnet must be
fixed to the cantilever.  Then, using the techniques of conventional
NMR described above, the nuclear spins are made to periodically flip,
generating an oscillating magnetic force acting on the cantilever.  In
order to resonantly excite the cantilever, the nuclear spins must be
inverted at the cantilever's mechanical resonance frequency.  The
cantilever's mechanical oscillations are then measured by an optical
interferometer or beam deflection detector. The electronic signal
produced by the optical detector is proportional both to the
cantilever oscillation amplitude and the number of nuclear spins in
the imaging volume.  Spatial resolution results from the fact that the
nano-magnet produces a magnetic field which is a strong function of
position.  The magnetic resonance condition and therefore the region
where the spins periodically flip is confined to a thin, approximately
hemispherical ``resonant slice'' that extends outward from the
nano-magnet (see Figs.~\ref{fig1} and \ref{fig6}).  By scanning the
sample in 3D through this resonant region, a spatial map of the
nuclear spin density can be made.

The advantage of force-detected over inductive techniques is that much
smaller devices can be made. In the latter case, the measurement can
only be sensitive if the nuclear spins significantly alter the
magnetic field within the pick-up coil, i.e.\ if the spins fill a
significant fraction of the coil volume. For spin ensembles with
volumes significantly smaller than (1 $\mu$m)$^3$, it is extremely
challenging to realize pick-up coils small enough to ensure an
adequate filling factor. As a result, even the best resolutions
achieved by inductively detected MRI require sample volumes of at
least (3 $\mu$m)$^3$ \cite{Ciobanu:2002}.  Mechanical resonators, in
contrast, can now be fabricated with dimensions far below a micron,
such that the sample's mass (which is the equivalent to the filling
volume in a pick-up coil) is always significant compared to the bare
resonator mass. In addition, mechanical devices usually show resonant
quality factors that surpass those of inductive circuits by orders of
magnitude, resulting in a much lower baseline noise. For example,
state-of-the art cantilever force transducers achieve quality factors
between $10^4$ and $10^7$, enabling the detection of forces of
aN/Hz$^{1/2}$ -- less than a billionth of the force needed to break a
single chemical bond. In addition, scanning probe microscopy offers
the stability to position and image samples with nanometer precision.
The combination of these features allows mechanically detected MRI to
image at resolutions that are far below one micrometer, and -- in
principle -- to aspire to atomic resolution.

\subsection{Early MRFM}
\label{EarlyMRFM}

The use of force-detection techniques in NMR experiments dates back to
Evans in 1955 \cite{Evans:1955}, and was also used in torque
magnetometry measurements by Alzetta and coworkers in the sixties
\cite{Alzetta:1967}.  In 1991 Sidles, independent of this very early
work, proposed that magnetic resonance detection and imaging with
atomic resolution could be achieved using microfabricated cantilevers
and nanoscale ferromagnets.  The first micrometer-scale experimental
demonstration using cantilevers was realized by Rugar
\cite{Rugar:1992}, demonstrating mechanically-detected electron spin
resonance in a 30-ng sample of diphenylpicrylhydrazil (DPPH).  The
original apparatus operated in vacuum and at room temperature with the
DPPH sample attached to the cantilever.  A mm-sized coil produced an
rf magnetic field tuned to the electron spin resonance of the DPPH
(220 MHz) with a magnitude of 1 mT.  By changing the strength of a
polarizing magnetic field (8 mT) in time, the electron spin
magnetization in the DPPH was modulated.  In a magnetic field gradient
of 60 T/m, produced by a nearby NdFeB permanent magnet, the sample's
oscillating magnetization resulted in a time-varying force between the
sample and the magnet.  This force modulation was converted into
mechanical vibration by the compliant cantilever.  Displacement
oscillations were detected by a fiber-optic interferometer achieving a
thermally limited force sensitivity of 3 fN/$\sqrt{\text{Hz}}$.

\begin{figure}\includegraphics[width=3.2in]{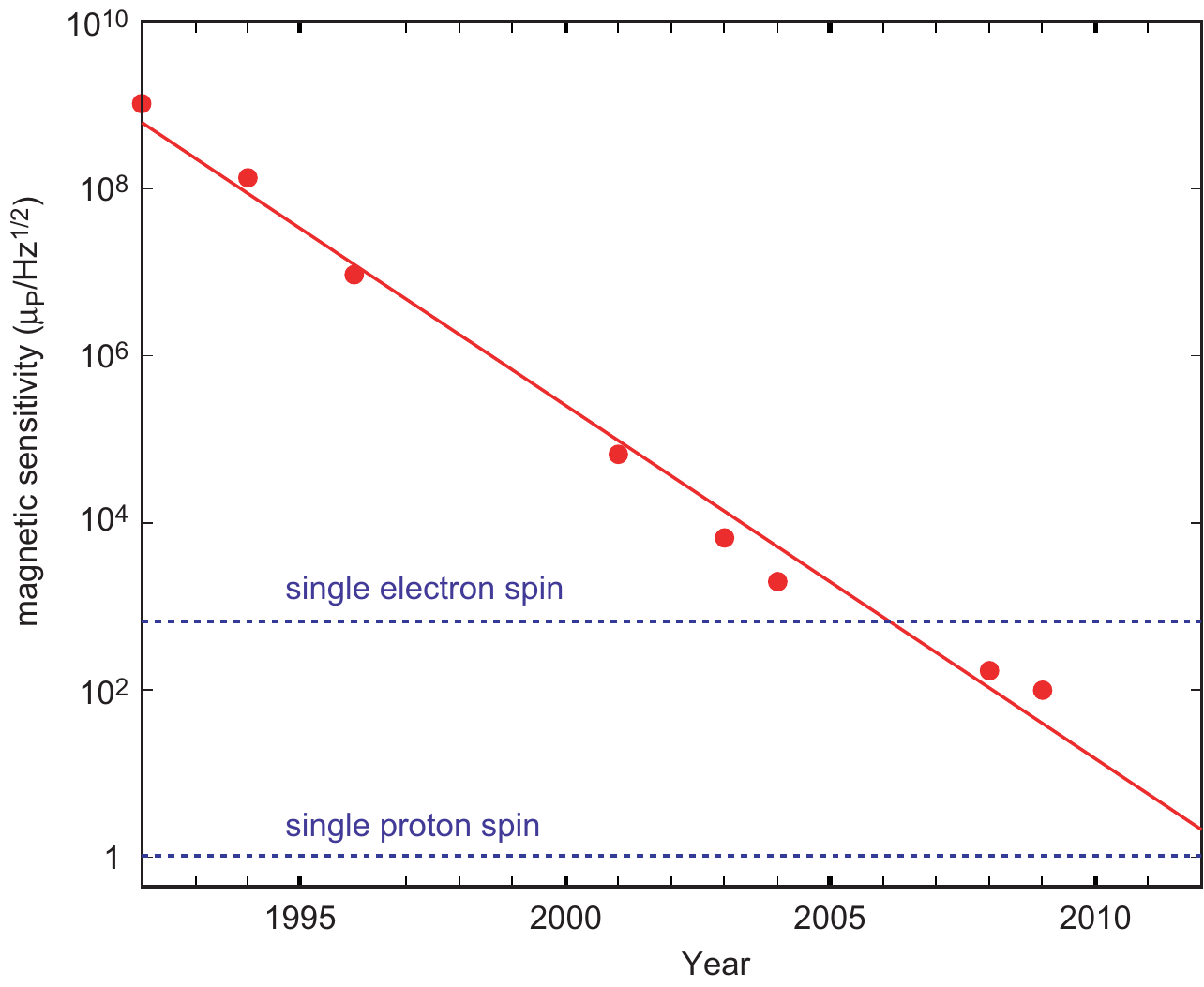}
  \caption{\footnotesize Advances in the sensitivity of force-detected
    magnetic resonance over time.  Remarkably, improvements have
    closely followed a ``Moore's law'' for over a decade, with the
    magnetic moment sensitivity doubling roughly every eight
    months. Dots are experimental values
    \cite{Rugar:1992,Rugar:1994,Wago:1996,Stipe:2001,Mamin:2003,Rugar:2004,DegenTMV:2009,Mamin:2009},
    and dashed lines indicate sensitivities of one electron and one
    proton magnetic moment ($\mu_{\rm P}$), respectively.}
\label{fig2}
\end{figure}

During the years following this initial demonstration of
cantilever-based magnetic resonance detection, the technique has
undergone a series of developments towards higher sensitives that, as
of today, is about $10^7$ times that of the 1992 experiment (see
Fig.~\ref{fig2}).  In the following, we briefly review the important
steps that led to these advances while also touching on the
application of the technique to imaging and magnetic resonance
spectroscopy.  Several review articles and book chapters have appeared
in the literature that discuss some of these earlier steps more
broadly and in richer detail
\cite{Sidles:1995,Nestle:2001,Suter:2004,Kuehn:2008,Berman:2006,Barbic:2009}.

Following the initial demonstration of mechanically-detected electron
paramagnetic resonance (EPR) \cite{Rugar:1992}, the MRFM technique was
soon extended to NMR in 1994 \cite{Rugar:1994} and to ferromagnetic
resonance in 1996 \cite{Zhang:1996}. A major step towards higher
sensitivity was made by incorporating the MRFM instrument into a
cryogenic apparatus in order to reduce the thermal force noise of the
cantilever. A first experiment carried out in 1996 at a temperature of
14 K achieved a force sensitivity of 80 aN/$\sqrt{\text{Hz}}$
\cite{Wago:1996}, a roughly 50-fold improvement compared to 1992
mostly due to the higher cantilever mechanical quality factor and the
reduced thermal noise achieved at low temperatures. In 1998,
researchers introduced the ``tip-on-cantilever'' scheme
\cite{Wago:1998} (shown in Fig.~\ref{fig1}(a)), where the roles of
gradient magnet and sample were interchanged. Using this approach,
field gradients of up to $2.5 \times 10^5$ T/m were obtained by using
a magnetized sphere of 3.4-$\mu$m diameter \cite{Bruland:1998}.  These
gradients are more than three orders of magnitude larger than those
achieved in the first MRFM experiment.  In parallel, a series of spin
detection protocols were also invented.  These protocols include the
detection of spin signals in the form of a shift in the cantilever
resonance frequency (rather than changes in its oscillation amplitude)
\cite{Stipe:2001Relax}, and a scheme that relies on detecting a
force-gradient, rather than the force itself \cite{Garner:2004}. In
2003, researchers approached the level of sensitivity necessary to
measure statistical fluctuations in small ensembles of electron spins,
a phenomenon that had previously only been observed with long
averaging times \cite{Mamin:2003}.  Further refinements finally led to
the demonstration of single electron spin detection in 2004 by the IBM
group \cite{Rugar:2004}, which we discuss separately below.

While the bulk of MRFM experiments address the improvement of
detection sensitivity and methodology, effort has also been devoted to
demonstrate the 3D imaging capacity of the instrument.  The first
one-dimensional MRFM image was made using EPR detection in 1993 and
soon after was extended to two and three dimensions
\cite{Zuger:1993,Zuger:1994,Zuger:1996}.  These experiments reached
about 1-$\mu$m axial and 5-$\mu$m lateral spatial resolution, which is
roughly on par with the best conventional EPR microscopy experiments
today \cite{Blank:2003}. In 2003, sub-micrometer resolution (170 nm in
one dimension) was demonstrated with NMR on optically pumped GaAs
\cite{Thurber:2003}.  In parallel, researchers started applying the
technique for the 3D imaging of biological samples, like the liposome,
at micrometer resolutions \cite{Tsuji:2004}.  Shortly thereafter, a
80-nm voxel size was achieved in an EPR experiment that introduced an
iterative 3D image reconstruction technique \cite{Chao:2004}.  The
one-dimensional imaging resolution of the single electron spin
experiment in 2004, finally, was about 25 nm \cite{Rugar:2004}.

The prospect of applying the MRFM technique to nanoscale spectroscopic
analysis has also led to efforts towards combination with pulsed NMR
and EPR techniques.  MRFM is ill suited to high resolution
spectroscopy as broadening of resonance lines by the strong field
gradient of the magnetic tip completely dominates any intrinsic
spectral features.  Nevertheless, a number of advances have been made.
In 1997, MRFM experiments carried out on phosphorus-doped silicon were
able to observe the hyperfine splitting in the EPR spectrum
\cite{Wago:1997}.  Roughly at the same time, a series of basic pulsed
magnetic resonance schemes were demonstrated to work well with MRFM,
including spin nutation, spin echo, and $T_1$ and $T_{1\rho}$
measurements \cite{Wago:1998Echo,Schaff:1997}. In 2002, researchers
applied nutation spectroscopy to quadupolar nuclei in order to extract
local information on the quadrupole interaction \cite{Verhagen:2002}.
This work was followed by a line of experiments that demonstrated
various forms of NMR spectroscopy and contrast, invoking dipolar
couplings \cite{Degen:2005}, cross polarization
\cite{Lin:2005,Eberhardt:2007}, chemical shifts \cite{Eberhardt:2008},
and multi-dimensional spectroscopy \cite{Eberhardt:2008}. Some
interesting variants of MRFM that operate in homogeneous magnetic
fields were also explored.  These techniques include measurement of
torque rather than force \cite{Alzetta:1967,Ascoli:1996} and the
so-called ``Boomerang'' experiment \cite{Leskowitz:1998,Madsen:2004}.

Finally, while not within the scope of this review, it is worth
mentioning that MRFM has also been successfully applied to a number of
ferromagnetic resonances studies, in particular for probing the
resonance structure of micron-sized magnetic disks.
\cite{Wigen:2006,Loubens:2007}.
\begin{figure}\includegraphics[width=3.2in]{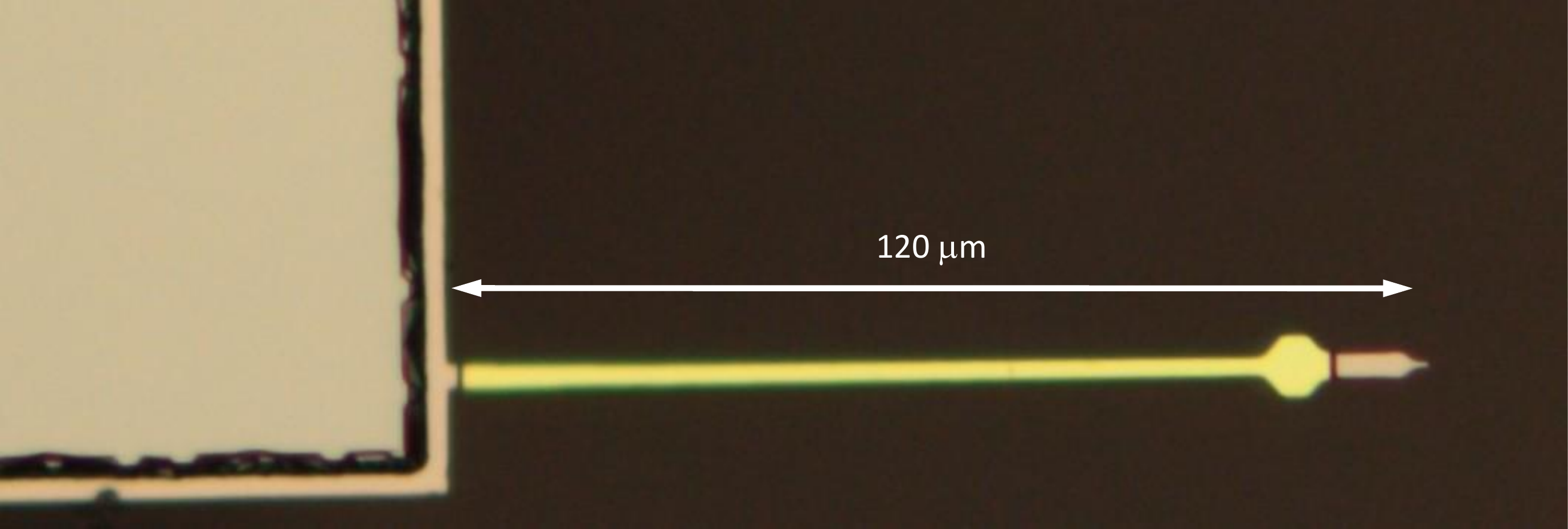}
  \caption{\footnotesize Image of an ultrasensitive mass-loaded Si
    cantilever taken from an optical microscope.  This type of
    cantilever, which is about 100-nm-thick and has a spring constant
    under 100 $\mu$N/m, has been used as a force transducer in the
    many of the latest MRFM experiments \cite{Chui:2003}.}
\label{fig3}
\end{figure}

\subsection{Single electron MRFM}
\label{singleEMRFM}

The measurement of a single electron spin by the IBM group in 2004
concluded a decade of development on the MRFM technique and stands out
as one of the first single-spin measurements in solid-state physics.
A variety of developments led to the exceptional measurement
sensitivity required for single-spin detection.  These include the
operation of the apparatus at cryogenic temperatures and high vacuum,
the ion-beam-milling of magnetic tips in order to produce large
gradients, and the fabrication of mass-loaded attonewton-sensitive
cantilevers \cite{Chui:2003} (shown in Fig.~\ref{fig3}).  The thermal
noise in higher order vibrational modes of mass-loaded cantilevers is
suppressed compared with the noise in the higher order modes of
conventional, ``flat'' cantilevers.  Since high frequency vibrational
noise in combination with a magnetic field gradient can disturb the
electron spin, the mass-loaded levers proved to be a crucial advance
for single-electron MRFM.  In addition, the IBM group developed a
sensitive interferometer employing only a few nanowatts of optical
power for the detection of cantilever displacement \cite{Mamin:2001}.
This low incident laser power is crucial for achieving low cantilever
temperatures and thus minimizing the effects of thermal force noise.
A low-background measurement protocol called OSCAR based on the NMR
technique of adiabatic rapid passage was also employed
\cite{Stipe:2001}.  Finally, the experiment required the construction
of an extremely stable measurement system capable of continuously
measuring for several days in an experiment whose single-shot
signal-to-noise ratio was just 0.06 \cite{Rugar:2004}.

The path to this experimental milestone led through a variety of
interesting physics experiments.  In 2003, for example, researchers
reported on the detection and manipulation of small ensembles of
electron spins -- ensembles so small that the their statistical
fluctuations dominate the polarization signal \cite{Mamin:2003}.  The
approach developed for measuring statistical polarizations provided a
potential solution to one of the fundamental challenges of performing
magnetic resonance experiments on small numbers of spins.  In 2005,
Budakian and coworkers took these concepts one step further by
actively modifying the statistics of the naturally occurring
fluctuations of spin polarization \cite{Budakian:2005}. In one
experiment, the researchers polarized the spin system by selectively
capturing the transient spin order. In a second experiment, they
demonstrated that spin fluctuations can be rectified through the
application of real-time feedback to the entire spin ensemble.

\section{Recent strides in nuclear MRFM}
\label{nMRFMResearch}

In the following, we summarize the latest advances made to nuclear
spin detection by MRFM. The shift of focus from electron to nuclear
spins is driven by the prospect of applying the technique for
high-resolution magnetic resonance microscopy. MRI has had a
revolutionary impact on the field of non-invasive medical screening,
and is finding an increased number of applications in materials
science and biology. The realization of MRI with nanometer or
sub-nanometer resolution may have a similar impact, for example, in
the field of structural biology. Using such a technique, it may be
possible to image complex biological structures, even down to the
scale of individual molecules, revealing features not elucidated by
other methods.

The detection of a single nuclear spin, however, is far more
challenging than that of single electron spin.  This is because the
magnetic moment of a nucleus is much smaller: a hydrogen nucleus
(proton), for example, possess a magnetic moment that is only $\sim
1/650$ of an electron spin moment. Other important nuclei, like
$^{13}$C or a variety of isotopes present in semiconductors, have even
weaker magnetic moments.  In order to observe single nuclear spins, it
is necessary to improve the state-of-the-art sensitivity by another
two to three orders of magnitude.  While not out of the question, this
is a daunting task that requires significant advances to all aspects
of the MRFM technique.  In the following, we discuss some steps made
in this direction since 2005.  Our focus is on the work contributed by
the authors while at the IBM Almaden Research Center.  There the
authors were part of a team led by Dan Rugar, who has pioneered many
of the important developments in MRFM since its experimental
beginnings in 1992.

\subsection{Improvements to microfabricated components}
\label{ExpImprovements}

Improvements in the sensitivity and resolution of mechanically
detected MRI hinge on a simple signal-to-noise ratio, which is given
by the ratio of the magnetic force power exerted on the cantilever
over the force noise power of the cantilever device.  For small
volumes of spins, we measure statistical spin polarizations, therefore
we are interested in force powers (or variances) rather than force
amplitudes:
\begin{equation}
\label{eq1}
\text{SNR} = N \frac{\left ( \mu_{N} G \right )^2}{S_{\rm F} B}.
\end{equation}
Here, $N$ is the number of spins in the detection volume, $\mu_{N}$ is
the magnetic moment of the nucleus of interest, $G$ is the magnetic
field gradient at the position of the sample, $S_{F}$ is the force
noise spectral density set by the fluctuations of the cantilever
sensor, and $B$ is the bandwidth of the measurement, determined by the
nuclear spin relaxation rate $1 / \tau_m$. This expression gives the
single-shot signal-to-noise ratio of a thermally-limited MRFM
apparatus. The larger this signal-to-noise ratio is, the better the
spin sensitivity will be.

From the four parameters appearing in (\ref{eq1}), only two can be
controlled and possibly improved. On the one hand, the magnetic field
gradient $G$ can be enhanced by using higher quality magnetic tips and
by bringing the sample closer to these tips. On the other hand, the
force noise spectral density $S_F$ can be reduced by going to lower
temperatures and by making intrinsically more sensitive mechanical
transducers.

\begin{figure}\includegraphics[width=3.2in]{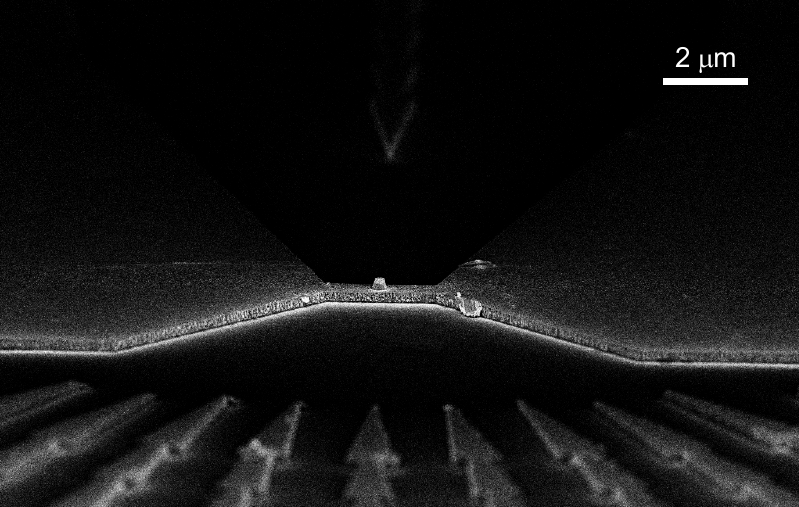}
\caption{\footnotesize
  An SEM image of a Cu ``microwire'' rf source with integrated FeCo
  tip for MRFM \cite{PoggioAPL:2007}. The arrow-like structures at the
  bottom of the image provide guidance for aligning the microwire with
  the cantilever.}
\label{fig4}
\end{figure}
The latest improvements to MRFM sensitivity rely on advances made to
both of these critical parameters. In 2006, the IBM group introduced a
micromachined array of Si cones as a template and deposited a
multilayer Fe/CoFe/Ru film to fabricate nanoscale magnetic tips
\cite{Mamin:2007}.  The micromachined tips produce magnetic field
gradients in excess of $10^6$ T/m owing to their sharpness (the tip
radius is less than 50 nm).  Previously, maximum gradients of $2
\times 10^5$ T/m had been achieved by ion-beam-milling SmCo particles
down to 150 nm in size.  The gradients from the new nanoscale tips
proved to be strong enough to push the resolution of MRI to below 100
nm, in an experiment that is further discussed in the next section.

In the two following years, the group made further improvements to
their measurement sensitivity through the development of a magnetic
tip integrated onto an efficient ``microwire'' rf source
\cite{PoggioAPL:2007}, illustrated in Fig.~\ref{fig4}. This change in
the apparatus solved a simple but significant problem: the typical
solenoid coils used to to generate the strong rf pulses for spin
manipulation dissipate large amounts of power, which even for very
small microcoils with a diameter of 300 $\mu$m amounts to over 0.2 W.
This large amount of heat is far greater than the cooling power of
available dilution refrigerators. As a result, nuclear spin MRFM
experiments had to be performed at elevated temperatures (4 K or
higher), thereby degrading the SNR.  In some cases the effects can be
mitigated through pulse protocols with reduced duty cycles
\cite{Garner:2004,Mamin:2007}, but it is desirable to avoid the
heating issue altogether.

Micro-striplines, on the other hand, can be made with sub-micrometer
dimensions using e-beam lithography techniques. Due to the small size,
the stripline confines the rf field to a much smaller volume and
causes minimal heat dissipation.  Using e-beam lithography and
lift-off, the IBM group fabricated a Cu ``microwire'' device that was
0.2 $\mu$m thick, 2.6 $\mu$m long, and 1.0 $\mu$m wide. A
stencil-based process was then used to deposit a 200-nm-diameter FeCo
tip on top of the wire to provide a static magnetic field gradient.
Since the sample could be placed within 100 nm of the microwire and
magnetic tip, rf magnetic fields of over 4 mT could be generated at
115 MHz with less than 350 $\mu$W of dissipated power. As a result,
the cantilever temperature during continuous rf irradiation could be
stabilized below 1 K, limited by other experimental factors and not
the rf device.  Simultaneously, the cylindrical geometry of the
magnetic tip optimized the lateral field gradient as compared to the
micromachined thin-film Si tips, resulting in values exceeding
$4\times10^6$ T/m. As an added benefit, the alignment of the apparatus
was simplified as the magnetic tip and the rf source were integrated
on a single chip.  The cantilever carrying the sample simply needed to
be positioned directly above the microwire device.  Previous
experiments had required an involved three part alignment of
magnetic-tipped cantilever, sample, and rf source.

\subsection{MRI with resolution better than 100 nm}
\label{100nm}

The above instrumental advances to the technique led to two
significant experiments that finally demonstrated MRFM imaging
resolutions in the low nano-scale. In a first experiment in 2007,
Mamin and coworkers used a ``sample-on-cantilever'' geometry with a
patterned 100-nm-thick CaF$_2$ film as their sample and a
micromachined Si tip array coated with a thin magnetic layer as their
magnetic tip \cite{Mamin:2007}. The CaF$_2$ films were thermally
evaporated onto the end of the cantilever and then patterned using a
focused ion beam, creating features with dimensions between 50 and 300
nm. The cantilevers used in these measurements were custom-made
single-crystal Si cantilevers with a 60 $\mu$N/m spring constant
\cite{Chui:2003}.

\begin{figure}\includegraphics[width=3.2in]{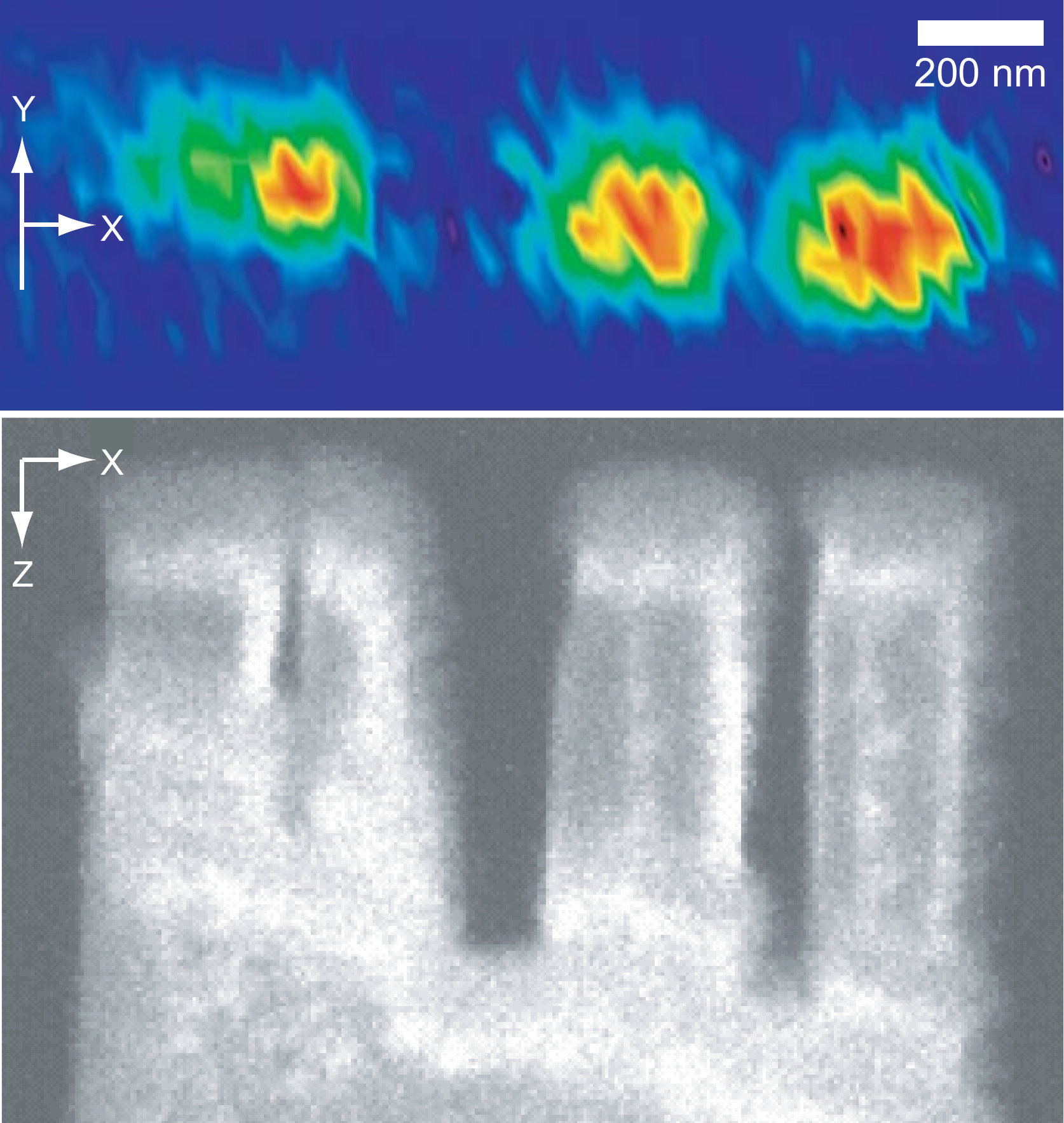}
\caption{\footnotesize
  (a) Two-dimensional MRFM image of $^{19}$F nuclear spins in a
  patterned CaF$_2$ sample, and (b) corrsponding SEM micrograph (side view) of the
  cantilever end with the 80 nm thin CaF$_2$ film at the top of the
  image. Figure adapted from Ref. \cite{Mamin:2007}. }
\label{fig5}
\end{figure}

Fig.~\ref{fig5} shows the result of such an imaging experiment,
measuring the $^{19}$F nuclei in the CaF$_2$ sample. The resultant
image reproduced the morphology of the CaF$_2$ sample, which consisted
of several islands of material, roughly 200-nm-wide and 80-nm-thick,
at a lateral resolution of 90 nm. At a temperature of 600 mK and after
10 minutes of averaging, the achieved detection sensitivity (SNR of
one) corresponded to the magnetization of about 1200 $^{19}$F nuclear
moments.

\subsection{Nanoscale MRI of virus particles}
\label{VirusImaging}

Following the introduction of the integrated microwire and tip device,
the IBM researchers were able to improve imaging resolutions to well
below 10 nm \cite{DegenTMV:2009}. These experiments, which used single
tobacco mosaic virus (TMV) particles as the sample, both show the
feasibility for MRI imaging with nanometer resolution, and the
applicability of MRFM to biologically relevant samples.
\begin{figure}\includegraphics[width=3.2in]{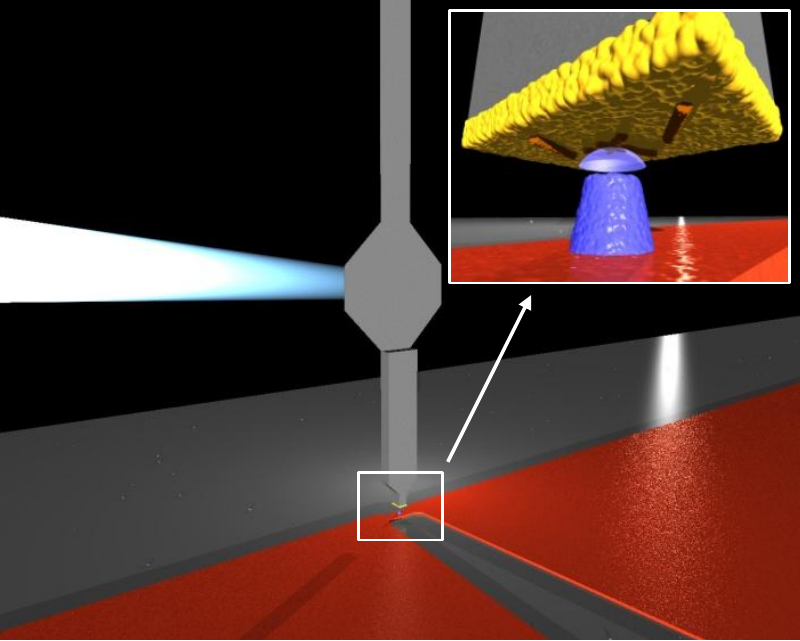}
  \caption{\footnotesize Artistic view of the MRFM apparatus used for
    MRI of individual tobacco mosaic virus particles.  Pictured in the
    center is the cantilever, coming from the left is the laser beam
    used for position sensing, and in red is the Cu microwire rf
    source.  The inset shows a close-up representation of the
    gold-coated end of the cantilever with attached virus particles.
    On top of the microwire, in blue, is the magnetic FeCo tip with
    the ``mushroom'' shaped resonant slice hovering above. }
\label{fig6}
\end{figure}

Fig.~\ref{fig6} is a representation of the MRFM apparatus used in
these experiments. The virus particles were transferred to the
cantilever end by dipping the tip of the cantilever into a droplet of
aqueous solution containing suspended TMV. As a result, some TMV were
attached to the gold layer previously deposited on the cantilever end,
The density of TMV on the gold layer was low enough that individual
particles could be isolated. Then the cantilever was mounted into the
low-temperature, ultra-high-vacuum measurement system and aligned over
the microwire.

After applying a static magnetic field of about 3 T, resonant rf
pulses were applied to the microwire source in order to flip the $^1$H
nuclear spins at the cantilever's mechanical resonance.  Finally, the
end of the cantilever was mechanically scanned in three dimensions
over the magnetic tip.  Given the extended geometry of the region in
which the resonant condition is met, i.e. the ``resonant slice'', a
spatial scan does not directly produce a map of the $^1$H distribution
in the sample. Instead, each data point in the scan contains force
signal from $^1$H spins at a variety of different positions.  In order
to reconstruct the three-dimensional spin density (the MRI image), the
force map must be deconvolved by the point spread function defined by
the resonant slice.  Fortunately, this point spread function can be
accurately determined using a magnetostatic model based on the
physical geometry of the magnetic tip and the tip magnetization.
Deconvolution of the force map into the three-dimensional $^1$H spin
density can be done in several different ways; for the results
presented in \cite{DegenTMV:2009} the authors applied the iterative
Landweber deconvolution procedure suggested in an earlier MRFM
experiment \cite{Chao:2004,Dobigeon:2009}.  This iterative algorithm
starts with an initial estimate for the spin density of the object and
then improves the estimate successively by minimizing the difference
between the measured and predicted spin signal maps.  The iterations
proceed until the residual error becomes comparable with the
measurement noise.

\begin{figure}[t]\includegraphics[width=3.2in]{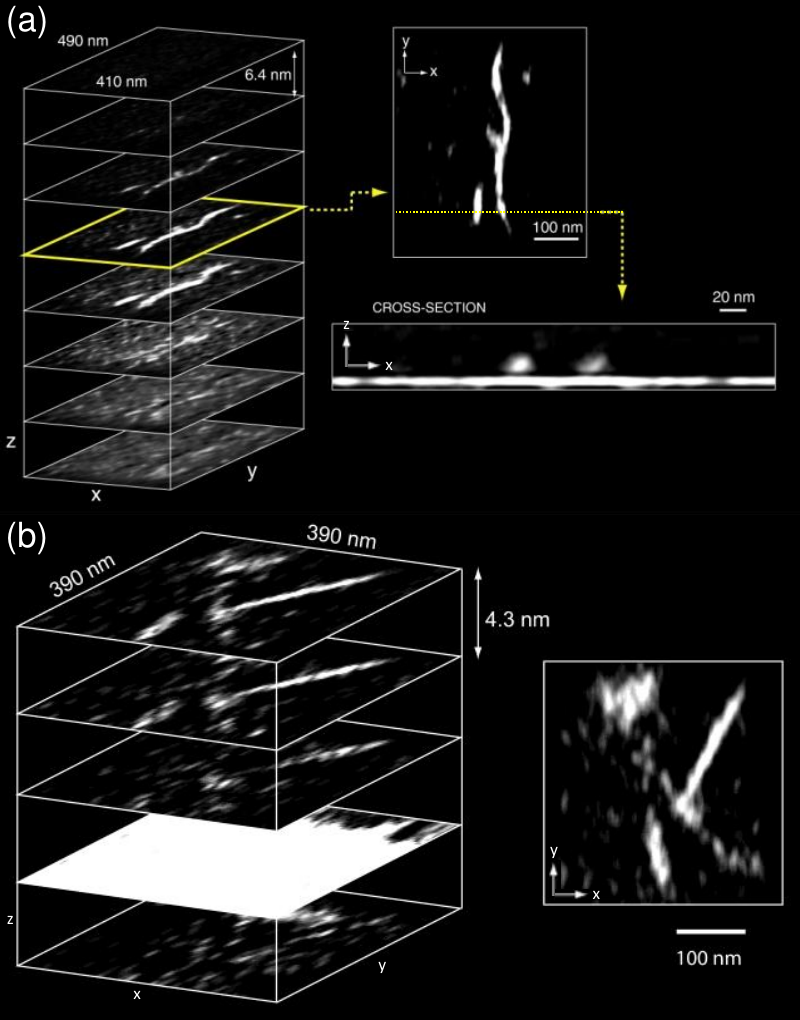}
\caption{\footnotesize
  Nanoscale MRI images of tobacco mosaic virus (TMV) particles
  acquired by MRFM.  (a) The series of images to the left depicts the
  3D $^1$H spin density of virus particles deposited on the end of the
  cantilever.  Black represents very low or zero density of hydrogen,
  while white is high hydrogen density.  The right side shows a
  representative xy-plane, with several viral fragments visible, and a
  cross-section (xz-plane) of two virus particles that reveals an
  underlying molecular layer of hydrocarbons covering the cantilever
  surface.  (b) 3D $^1$H spin density recorded on a different region
  of the same cantilever as in (a), showing an intact and several
  fragmented virus particles.  The right side shows a representative
  xy-plane.}
\label{fig7}
\end{figure}

\begin{figure}[t]\includegraphics[width=3.2in]{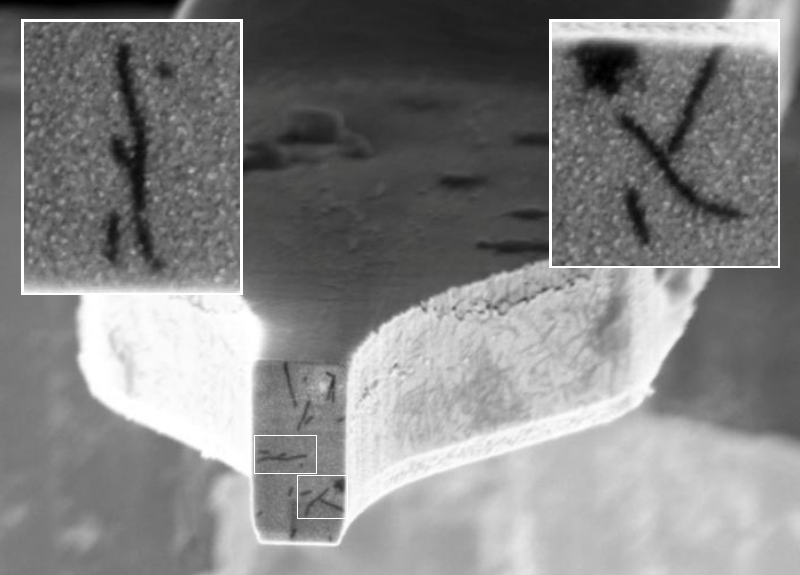}
  \caption{\footnotesize SEM of the TMV particles and particle
    fragments on the gold-coated cantilever end.  The insets enlarge
    the areas that were imaged by MRFM in Fig.~\ref{fig7}. }
\label{fig8}
\end{figure}

The result of a representative experiment is shown in Fig.~\ref{fig7}.
Here, clear features of individual TMV particles, which are
cylindrical, roughly 300-nm-long, and 18 nm in diameter, are visible
and can be confirmed against a scanning electron micrograph (SEM) of
the same region (Fig.~\ref{fig8}).  As is often the case, both whole
virus particles and particle fragments are observed.  Note that the
origin of contrast in MRFM image and the SEM image is very different:
the MRFM reconstruction is elementally specific and shows the 3D
distribution of hydrogen in the sample; contrast in the SEM image is
mainly due to the virus blocking secondary electrons emitted from the
underlying gold-coated cantilever surface.  In fact, the SEM image had
to be taken after the MRFM image as exposure to the electron beam
destroys the virus particles.  The imaging resolution, while not fine
enough to discern any internal structure of the virus particles,
constitutes a 1000-fold improvement over conventional MRI, and a
corresponding improvement of volume sensitivity by about 100 million.

\subsection{Imaging organic nanolayers}
\label{NanoLayers}

In addition to ``seeing'' individual viruses, the researchers also
detected an underlying proton-rich layer.  This signal originated from
a naturally occurring, sub-nanometer thick layer of adsorbed water
and/or hydrocarbon contamination.

The hydrogen-containing adsorbates picked up on a freshly cleaned gold
surface turn out to be enough to produce a distinguishable and
characteristic signal.  From analysis of the signal magnitude and
magnetic field dependence, the scientists were able to determine that
the adsorbates form a uniform layer on the gold surface with a
thickness of roughly 5 to 10 \AA \cite{Mamin:2009}.

\begin{figure}[b]\includegraphics[width=3.2in]{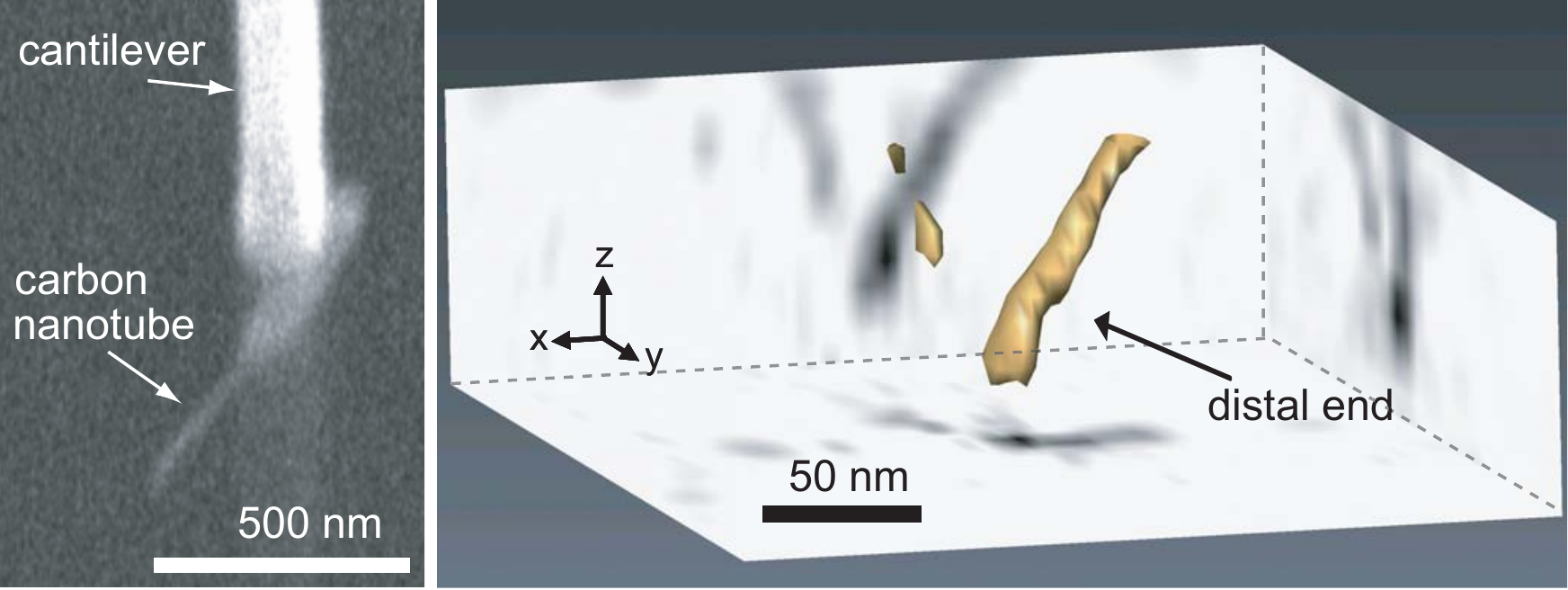}
\caption{\footnotesize
  Scanning electron microscopy image of a 10-nm-diameter carbon
  nanotube attached to the end of a Si cantilever (left), and MRFM
  image of the proton density at the nanotube's distal end (right).
  Figure adapted from Ref. \cite{Mamin:2009}.}
\label{fig9}
\end{figure}

Using a similar approach, the researchers made a 3D image of a
multiwalled nanotube roughly 10 nm in diameter, depicted in
Fig.~\ref{fig9}.  The nanotube, attached to the end of a 100-nm-thick
Si cantilever, protruded a few hundred nanometers from the end of the
cantilever.  As had been previously observed with gold layers, the
nanotube was covered by a naturally occurring proton-containing
contamination layer.  Though the magnitude of the signal was roughly
10 times less than that of the two-dimensional layer -- reflecting its
relatively small volume -- it was accompanied by a very low background
noise level that made it possible to produce a clear image of the
morphology of the nanotube.  Using the same iterative deconvolution
scheme developed to reconstruct the image of the TMV particles, the
researchers produced an image of a cylindrical object, 10 nm in
diameter at the distal end.  No evidence was found for the hollow
structure that might be expected in the image of such a layer.  The
experiment did not show any evidence for an empty cylindrical region
within the nanotube. Given the small inner diameter (less than 10 nm),
however, it was not clear whether hydrogen-containing material was in
fact incorporated inside the nanotube, or if the resolution of the
image was simply not sufficient to resolve the feature.

\subsection{Observation and manipulation of statistical polarizations}
\label{StatPolarization}

\begin{figure}\includegraphics[width=3.2in]{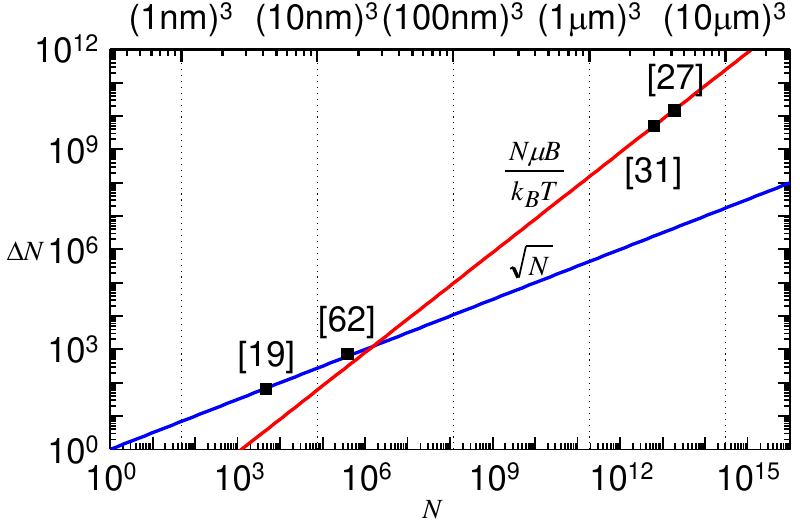}
\caption{\footnotesize
  Comparison of mean thermal (or Boltzmann) polarization $\bar{\Delta
    N} = N\mu B/k_{\rm B}T$ versus statistical polarization $\Delta
  N_{\rm rms}=\sqrt{N}$ as a function of the number $N$ of nuclear
  spins in the ensemble.  While statistical polarization fluctuations
  are negligible for macroscopic samples where $N$ is large, they
  dominate over thermal polarization for small $N$.  Under the
  conditions typical for MRFM, where $T = 4$ K and $B = 3$ T (shown
  here), this crossover occurs around $N\approx10^6$, or for sample
  volumes below about (100 nm)$^3$.  Dots represent experimental
  values for conventional MRI \cite{Ciobanu:2002} and MRFM
  \cite{DegenTMV:2009,Mamin:2007,Rugar:1994}.  }
\label{fig10}
\end{figure}

While predominantly driven by the interest in high resolution MRI
microscopy, the exquisite spin sensitivity of MRFM also gives us a
window into the spin dynamics of small ensembles of spins. When
probing nuclear spins on the nanometer scale, for example, random
fluctuations of the spin polarization will typically exceed the mean
Boltzmann polarization if sample volumes are smaller than about (100
nm)$^3$, as shown in Fig.~\ref{fig10}.  This statistical polarization
arises from the incomplete cancellation of randomly oriented spins.
For an ensemble of $N$ nuclei of spin 1/2 and in the limit of small
mean polarization, which is representative of most experiments, the
variance of the fluctuations is $\sigma^2_{\Delta N} \simeq N$.  The
existence of statistical polarization was pointed out by Bloch in his
seminal paper on nuclear induction \cite{Bloch:1946}, and has been
observed experimentally by a number of techniques, including
superconducting quantum interference devices \cite{Sleator:1985},
conventional magnetic resonance detection
\cite{Mccoy:1989,Gueron:1989,Muller:2006}, optical techniques
\cite{Crooker:2004}, and MRFM \cite{Mamin:2005,Mamin:2007}.

In a result that was enabled by the latest advances in MRFM detection
sensitivity, the IBM scientists were able -- for the first time -- to
follow the fluctuations of a statistical polarization of nuclear spins
in real time.  These experiments followed the dynamics of an ensemble
of roughly $2 \times 10^6$ $^{19}$F spins in CaF$_2$
\cite{Degen:2007}.  The challenge of measuring statistical
fluctuations presents a major obstacle to nanoscale imaging
experiments.  In particular, the statistical polarization has random
sign and a fluctuating magnitude, making it hard to average signals.
An efficient strategy for imaging spin fluctuations is therefore to
use polarization variance, rather than the polarization itself, as the
image signal.  This has recently been demonstrated both by
force-detected \cite{Mamin:2007,Degen:2007,DegenTMV:2009,Mamin:2009}
and conventional \cite{Muller:2006} MRI.  Furthermore, it was
demonstrated that for cases where spin lifetimes are long, rapid
randomization of the spins by rf pulses can considerably enhance the
signal-to-noise ratio of the image \cite{Degen:2007}.  In the end, for
the purposes of imaging, it is not necessary to follow the sign of the
spin polarization; it is enough to simply determine from the measured
spin noise where and how many spins are present at a particular
location.

The nuclear spin lifetime itself, which is apparent as the correlation
time of the nuclear fluctuations $\tau_m$, was also shown to be an
important source of information.  Using suitable rf pulses,
researchers demonstrated that Rabi nutations, rotating-frame
relaxation times, and nuclear cross-polarization can be encoded in
$\tau_m$ leading to new forms of image contrast
\cite{PoggioAPL:2007,Poggio:2009}.

\subsection{Mechanically induced spin relaxation}
\label{MechRelax}

The high sensitivity of MRFM is enabled in part by the strong coupling
that can be achieved between spins and the cantilever.  This coupling
is mediated by field gradients that can exceed $5 \times 10^6$ T/m.
The strong interaction between spins and sensor has been the subject
of a number of theoretical studies, and is predicted to lead to a host
of intriguing effects.  These range from shortening of spin lifetimes
by ``back action'' \cite{Mozyrsky:2003,Berman:2003}, to spin alignment
by specific mechanical modes either at the Larmor frequency or in the
rotating frame \cite{Magusin:2000,Butler:2005}, to resonant
amplification of mechanical oscillations \cite{Bargatin:2003}, to
long-range mediation of spin couplings using charged resonator arrays
\cite{Rabl:2009}.

\begin{figure}\includegraphics[width=3.2in]{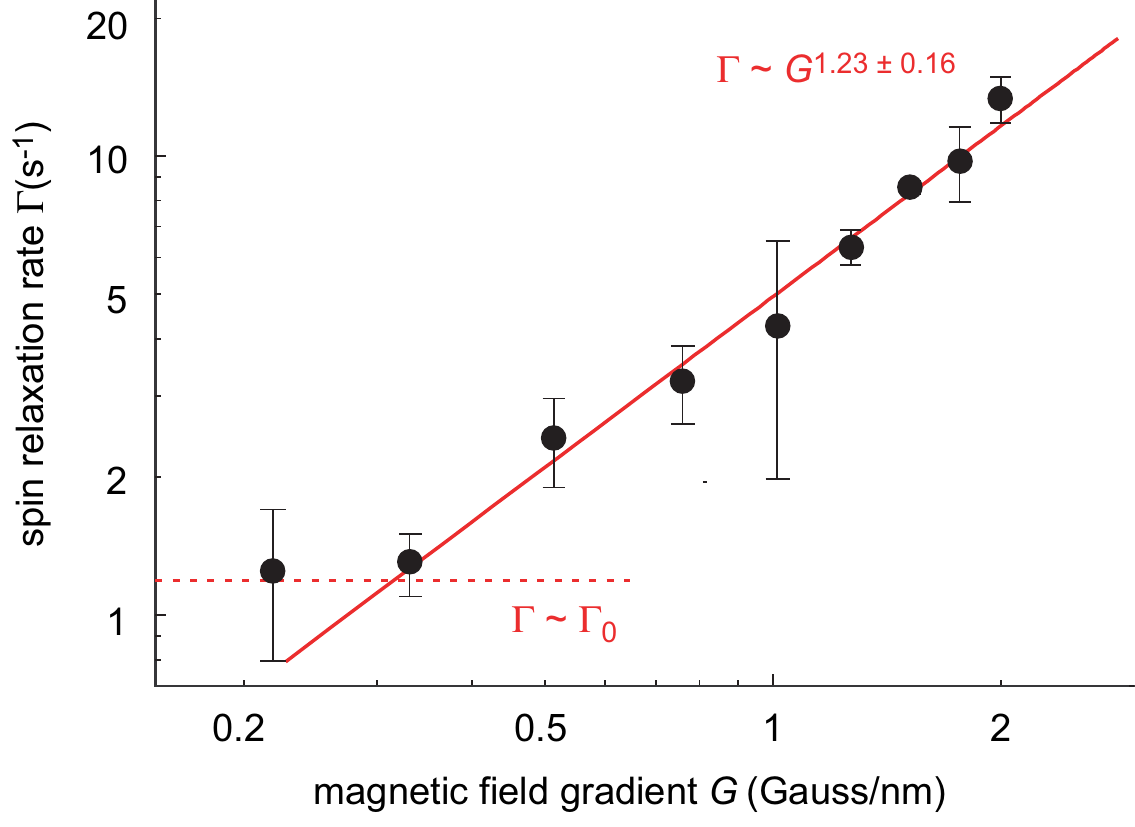}
\caption{\footnotesize
  Spin relaxation rate $\Gamma$ as a function of magnetic field
  gradient $G$.  In the weak coupling regime, nuclear spin relaxation
  is dominated by interaction with lattice phonons
  ($\Gamma=\Gamma_0$).  In the strong coupling regime, spins relax via
  a specific low-frequency mechanical mode of the cantilever and
  $\Gamma\propto G^{-1.23}$.  Figure adapted from Ref.
  \cite{DegenPRL:2008}.}
\label{fig11}
\end{figure}

Recently the IBM group reported the first direct experimental evidence
for accelerated nuclear spin relaxation induced by a single,
low-frequency mechanical mode \cite{DegenPRL:2008}.  In these
experiments the slight thermal vibration of the cantilever generated
enough magnetic noise to destabilize the spin.  Enhanced relaxation
was found when one of the cantilever's upper modes (in particular the
third mode with a frequency of about 120 kHz) coincided with the Rabi
frequency of the spins.  In this ``strong coupling'' regime, the
cantilever is more tightly coupled to one mechanical resonator mode
than to the continuum of phonons that are normally responsible for
spin-lattice relaxtation. Interestingly, these initial experiments
showed a scaling behavior of the spin relaxation rate with important
parameters, including magnetic field gradient and temperature, that is
substantially smaller than predicted by theory (see Fig.~\ref{fig11}).

\subsection{Force detected nuclear double resonance}
\label{DoubleResonance}

Most recently, the IBM group exploited couplings between different
spin species to enhance the 3D imaging capability of MRFM with the
chemical selectivity intrinsic to magnetic resonance.  They developed
a method of nuclear double-resonance that allows the enhancement of
the polarization fluctuation rate of one spin species by applying an
rf field to the second spin species, resulting in suppression of the
MRFM signal \cite{Poggio:2009}.  The physics behind this approach is
analogous to Hartmann-Hahn cross-polarization (CP) in NMR spectroscopy
\cite{Hartmann:1962}, but involves statistical rather than Boltzmann
polarization.  The IBM group was inspired by previous work done with
Boltzmann polarizations at the ETH in Z\"{u}rich demonstrating CP as
an efficient chemical contrast mechanism for micrometer-scale
one-dimensional MRFM imaging
\cite{Lin:2006,Eberhardt:2007,Eberhardt:2008}.  In the IBM experiment,
MRFM was used to measure the transfer between statistical
polarizations of $^1$H and $^{13}$C spins in $^{13}$C-enriched stearic
acid.  The development of a cross-polarization technique for
statistical ensembles adds an important tool for generating chemical
contrast to the recently demonstrated technique of nanometer-scale
MRI.

\section{Future developments}
\label{Developments}

Since its invention and early experimental demonstration in the
nineties \cite{Sidles:1991,Rugar:1992,Rugar:1994}, the MRFM technique
has progressed in its magnetic sensitivity from the equivalent of
$10^9$ to presently about 100 proton magnetic moments (see
Fig.~\ref{fig3}). In order to eventually detect single nuclear spins
and to image molecules at atomic resolution, the signal-to-noise ratio
of the measurement must still improve by two orders of magnitude.  It
is not clear if these advances can be achieved by incremental progress
to the key components of the instrument, i.e.\ cantilever force
transducers and nanoscale magnetic tips, or whether major shifts in
instrumentation and methodology will be necessary. In the following we
review some of the key issues and potential avenues for future
developments.

\subsection{Magnetic tips}

The magnetic force on the cantilever can be enhanced by increasing the
magnetic field gradient $G$. This can be achieved by making higher
quality magnetic tips with sharp features and high-moment materials,
and by simultaneously bringing the sample closer to these tips.  To
date, the highest magnetic field gradients have been reported in
studies of magnetic disk drive heads, ranging between $20\times10^6$
and $40\times10^6$ T/m \cite{Tsang:2006}. The pole tips used in drive
heads are typically made of soft, high-moment materials like FeCo, and
have widths below 100 nm. The magnetic tips used in the latest MRFM
experiments, on the other hand, are more than 200 nm in diameter, and
generate field gradients of less than $5\times10^6$ T/m. Moreover,
calculations indicate that these tips do not achieve the ideal
gradients which one would calculate assuming that they were made of
pure magnetic material.  This discrepancy may be due to a dead-layer
on the outside of the tips, to defects inside the tips, or to
contamination of the magnetic material.  By improving the material
properties and shrinking dimension of present MRFM tips, $G$ could be
increased by up to a factor of ten.  In practice, however, it will be
difficult to gain an order of magnitude in signal-to-noise purely by
improving the magnetic tips.  To achieve higher gradients -- and
therefore higher signal-to-noise -- we must resort to decreasing the
tip-sample spacing.

\subsection{Force noise near surfaces}

Since the gradient strength falls off rapidly with distance, bringing
the sample closer to the magnetic tip would also increase the field
gradient. However, measurements at small tip-sample spacings are
hampered by strong tip-sample interactions which produce mechanical
noise and dissipation in the cantilever. At the moment, imaging
experiments are limited to spacings on the order of 25 nm.  For some
experimental arrangements, surface dissipation can be observed at
separations well over 100 nm.  This interaction has been studied in
similar systems \cite{StipeFriction:2001,Kuehn:2006} and several
mechanisms have been proposed to explain its origin depending on the
details of the configuration
\cite{Persson:1998,Zurita:2004,Volokitin:2003,Volokitin:2005,Labaziewicz:2008}.
Most explanations point to trapped charges or dielectric losses in
either the substrate or the cantilever.  Experimentally, several
strategies could mitigate non-contact friction effects, including
chemical modification of the surface, narrow tip size, or
high-frequency operation.  None of these approaches has yet emerged as
the clear path for future improvement.

\subsection{Mechanical transducers}

The second means to improving the signal-to-noise ratio is the
development of more sensitive mechanical transducers, i.e.\ 
transducers that exhibit a lower force noise spectral density $S_F$.
For a mechanical resonator, $S_F$ is given by:
\begin{equation}
\label{eq2}
S_{F} = \frac{4 k_B T m \omega_0}{Q},
\end{equation}
where $k_B$ is the Boltzmann constant, $T$ is the temperature, $m$ is
the effective motional mass of the cantilever, $\omega_0$ is the
angular resonance frequency of the cantilever's fundamental mode, and
$Q$ the mechanical quality factor. In practice, this requires going to
lower temperatures and making cantilevers which simultaneously have
low $m \omega_0$ and large $Q$.

None of these steps is as straightforward as it first appears.
Temperatures below 100 mK can be achieved by cooling the apparatus in
a dilution refrigerator. In the latest experiments, however, the
cantilever temperature was limited to about 1 K by laser heating from
the interferometer displacement sensor, not by the base temperature of
the apparatus. Progress to sub-100 mK temperatures will therefore
require new developments in displacement sensing.

The best strategies for maximizing $Q$ are not well-understood
either. Apart from a loose trend that $Q$ often scales with thickness
\cite{Yasumura:2000}, and a few general rules of thumb, i.e.\
minimizing clamping losses by design and keeping the mechanical
resonator pristine and free of defects and impurities, no clear path
has emerged. Holding $Q$ constant, one finds from simple
Euler-Bernoulli beam theory that the product of $m$ and $\omega_0$ is
minimized for cantilevers that are long and thin.

On a more fundamental level, it is worth considering the use of
different materials and alternative geometries.  Over the past few
years a variety of nanomechanical resonators have been developed which
rival the force sensitivities of the single crystal Si cantilevers
used in most MRFM experiments. Some examples are the SiN membranes
serving as sample stages in transmission electron microscopy
\cite{Thompson:2008}, vapor grown silicon nanowires
\cite{Nichol:2008}, and strained SiN or aluminum beams
\cite{Teufel:2009,Rocheleau:2010}. With some exceptions, the general
trend is towards smaller resonators that more closely match the atomic
lengthscales of spins and molecules. Therefore, it appears likely that
future transducers will emerge as ``bottom-up'' structures rather than
the ``top-down'' structures of the past.  Instead of processing and
etching out small mechanical devices out of larger bulk crystals,
future resonators will probably be chemically grown or self-assembled:
For example, they will be macroscale ``molecules'' such as nanowires,
nanotubes \cite{Sazonova:2004}, or single sheets of graphene
\cite{Bunch:2007}.

Although uncontrolled bottom-up approaches tend to be ``dirty'',
remarkable mechanical properties can be achieved if care is taken to
keep this self assembly process ``clean''. Most recently, researchers
have demonstrated suspended carbon nanotubes with resonant frequencies
of 250 MHz, masses of $10^{-20}$ kg, and quality factors of $10^5$
\cite{Huttel:2009,Steele:2009}.  If such a carbon nanotube force
transducer could be operated at the thermal limit, which would require
improved displacement detectors capable of measuring the nanotube's
thermal motion, the resulting force sensitivity would be 0.01
aN/$\sqrt{\text{Hz}}$, about 50 times better than any known mechanical
force sensor today.

\subsection{Displacement sensors}

The mechanical deflection caused by spin or thermal force is typically
a fraction of an Angstrom. In order to transfer the deflection into
experimentally accessible electronic signals, very sensitive
displacement sensor must be employed.  To the best of our knowledge,
all MRFM measurements have made use of optical detectors based on
either optical beam deflection or laser interferometry.  While optical
methods provide an extremely sensitive means of measuring cantilever
displacement, they face limitations as cantilevers become smaller and
temperatures lower.

The first limitation comes about as the push for better spin
sensitivity necessitates smaller and smaller cantilevers.  The
reflective areas of these levers will shrink to the order of, or even
below, the wavelength of light.  As a result, optical sensors will
become less and less efficient as smaller and smaller fractions of the
incident light are reflected back from the resonators.  Thus, for the
next generation of cantilevers -- made from nanowires and nanotubes --
interferometric displacement sensing may no longer be an option.

In principle, the inefficient reflection from small resonators can be
balanced by increased laser power. Indeed, in a recent experiment,
Nichol and coworkers have been able to sense the motion of Si
nanowires at room temperature with diameters on the order of 20-nm
using optical interferometry \cite{Nichol:2008}.  The researchers used
a polarization resolved interferometer and a high incident laser power
in order to sense the cantilever's motion.

High optical powers are, however, not compatible with low temperature
operation.  Especially at millikelvin temperatures, most materials
(except for metals) have very poor thermal conductivities, and even
very low incident laser powers can heat the cantilever.  For example,
a laser power of only 20 nW from a 1550-nm laser is sufficient to
increase the temperature of a single crystal Si cantilever of the type
shown in Fig.~\ref{fig3} from less than 100 mK to 300 mK, even though
absorption is known to be minimal for this wavelength.

There are several potential displacement detectors which could achieve
better sensitivity than optical methods while causing less measurement
back-action, or heating.  An idea pursued by one of the authors is to
make an off-board capacitively-coupled cantilever displacement
detector based on a quantum point contact (QPC) \cite{Poggio:2008}.
Preliminary measurements indicate that such a detector reaches at
least the sensitivity of optical methods for equivalent cantilevers,
with no indication of back-action from the electrons flowing in the
device. While more work needs to be done, these kinds of capacitively
coupled detectors are promising means of measuring mechanical
resonators much smaller than the wavelength of light.  One might
imagine a future MRFM detection set-up where an arbitrarily small
cantilever could be used, and a capacitive displacement detector is
integrated on chip with a high-gradient magnetic tip, and an rf
microwire source.  Outstanding displacement sensitivities have also
been achieved with microwave interferometers
\cite{Teufel:2009,Rocheleau:2010}, superconducting single-electron
transistors, or high-finesse optical cavities made from micro-toroids
which are very sensitive to fluctuations of nearby objects
\cite{Anetsberger:2009}.  All of these latter displacement sensors
will, however, need adjustments in order to be integrated in a
contemporary scanning MRFM instrument.

\section{Comparison to other techniques}
\label{Othertechniques}

The unique position of MRFM among high-resolution microscopies becomes
apparent when comparing it to other, more established nanoscale
imaging techniques.  As a genuine scanning probe method, MRFM has the
potential to image matter at atomic resolution.  While atomic-scale
imaging is routinely achieved in scanning tunneling microscopy and
atomic force microscopy, these techniques are confined to the top
layer of atoms and cannot penetrate below surfaces
\cite{Hansma:1987,Giessibl:2003}.  Moreover, in standard scanning
probe microscopy (SPM), it is difficult and in many situations
impossible to identify the chemical species being imaged.  Since MRFM
combines SPM and MRI, these restrictions are lifted.  The
three-dimensional nature of MRI permits acquisition of sub-surface
images with high spatial resolution even if the probe is relatively
far away.  As with other magnetic resonance techniques, MRFM comes
with intrinsic elemental contrast and can draw from established NMR
spectroscopy procedures to perform detailed chemical analysis.  In
addition, MRI does not cause any radiation damage to samples, as do
electron and X-ray microscopies.

MRFM also distinguishes itself from super-resolution optical
microscopies that rely on fluorescence imaging \cite{Huang:2009}.  On
the one side, optical methods have the advantage of working \textit{in
  vivo} and they have the ability to selectively target the desired
parts of a cell.  Fluorescent labeling is now a mature technique which
is routinely used for cellular imaging.  On the other side, pushing
the resolution into the nanometer range is hampered by fundamental
limitations, in particular the high optical powers required and the
stability of the fluorophores.  Moreover, fluorescent labeling is
inextricably linked with a modification of the target biomolecules,
which alters the biofunctionality and limits imaging resolution to the
physical size of the fluorophores.

MRFM occupies a unique position among other nanoscale spin detection
approaches.  While single electron spin detection in solids has been
shown using several techniques, these mostly rely on the indirect
read-out via electronic charge \cite{Elzerman:2004,Xiao:2004} or
optical transitions \cite{Wrachtrup:1993,Jelezko:2002}.  In another
approach, the magnetic orientation of single atoms has been measured
via the spin-polarized current of a magnetic STM tip or using magnetic
exchange force microscopy \cite{Heinze:2000,Durkan:2002,Kaiser:2007}.
These tools are very valuable to study single surface atoms, however,
they are ill suited to map out sub-surface spins such as paramagnetic
defects.  In contrast, MRFM \textit{directly} measures the magnetic
moment of a spin, without resorting to other degrees of freedom,
making it a very general method.  This direct measurement of magnetic
moment (or magnetic stray field) could also be envisioned using other
techniques, namely SQuID microscopy \cite{Kirtley:1995}, Hall
microscopy \cite{Chang:1992}, or recently introduced diamond
magnetometry based on single nitrogen-vacancy centers
\cite{Degen:2008,Maze:2008,Balasubramanian:2008}.  So far, however,
none of these methods have reached the level of sensitivity needed to
detect single electron spins, or volumes of nuclear spins much less
than one micrometer \cite{DegenNNano:2008,Blank:2009}.  It is
certainly possible that future improvements to these methods --
especially to diamond magnetometry -- may result in alternative
techniques for nanoscale MRI that surpass the capabilities of MRFM.

\section{Outlook}
\label{Outlook}

Despite the tremendous improvements made to MRFM over the last decade,
several important obstacles must be overcome in order to turn the
technique into a useful tool for biologists and materials scientists.
Most existing MRFM instruments are technically involved prototypes;
major hardware simplifications will be required for routine screening
of nanoscale samples.  Suitable specimen preparation methods must be
developed that are compatible with the low temperature, high vacuum
environment required for the microscope to operate at its highest
sensitivity and resolution.  While this is particularly challenging
for biological samples, protocols exist which could be adapted to
MRFM.  In cryo-electron microscopy, for example, dispersed samples are
vitrified to preserve their native structure by plunge-freezing in
liquid nitrogen \cite{Taylor:1974}.  As objects become smaller,
isolation of samples and suppression of unwanted background signals
from surrounding material will become increasingly important.

The conditions under which the latest MRFM imaging experiments were
carried out are remarkably similar to those prevailing in
cryo-electron microscopy, the highest resolution 3D imaging technique
commonly used by structural biologists today.  Cryo-electron
microscopy, like MRFM, operates at low temperatures and in high
vacuum, requires long averaging times (on the order of days) to
achieve sufficient contrast, and routinely achieves resolutions of a
few nanometers \cite{Lucic:2005,Subramaniam:2005}.  Unlike MRFM,
however, electron microscopy suffers from fundamental limitations that
severely restrict its applicability.  Specimen damage by the
high-energy electron radiation limits resolution to 5-10 nm if only a
single copy of an object is available.  Averaging over hundreds to
thousands of copies is needed to achieve resolutions approaching 10
\AA \cite{Glaeser:2008}.  In addition, unstained images have
intrinsically low contrast, whereas staining comes at the expense of
modifying the native structure.  

MRFM has the unique capability to image nanoscale objects in a
non-invasive manner and to do so with intrinsic chemical selectivity.
For this reason the technique has the potential to extend microscopy
to the large class of structures that show disorder and therefore
cannot be averaged over many copies.  These structures include such
prominent examples as HIV, Influenza virus, and Amyloid fibrils.
Virtually all of these complexes are associated with important
biological functions ranging from a variety of diseases to the most
basic tasks within the cellular machinery.  For such complexes, MRFM
has the potential not only to image the three-dimensional
macromolecular arrangement, but also to selectively image specific
domains in the interior through isotopic labeling.

While the most exciting prospect for MRFM remains its application to
structural imaging in molecular biology, its applications are not
limited to biological matter.  For example, most semiconductors
contain non-zero nuclear magnetic moments.  Therefore MRFM may prove
useful for sub-surface imaging of nanoscale electronic devices.  MRFM
also appears to be the only technique capable of directly measuring
the dynamics of the small ensembles of nuclear spin that limit
electron spin coherence in single semiconductor quantum dots.  Polymer
films and self-assembled monolayers -- important for future molecular
electronics -- are another exciting target for MRFM and its capability
to image chemical composition on the nanoscale.  Finally, isotopically
engineered materials are becoming increasingly important for tuning a
variety of physical properties such as transport and spin.
Researchers currently lack a general method for non-invasively imaging
the isotopic composition of these materials
\cite{Shimizu:2006,Shlimak:2001,Kelly:2007}; MRFM techniques could
fill this void.

As force-detected magnetic resonance has traditionally been an
exploratory field, it is possible that applications other than
nanoscale imaging will emerge.  Single electron spin detection, for
example, is an important prerequisite for future quantum information
applications \cite{Rugar:2004,Kane:2000}.  At the same time, MRFM may
also become an important tool in the study of defects or dopants deep
in materials, or for mapping of spin labels in decorated biological
nanostructures \cite{Moore:2009}.  The key components to the
instrument -- in particular the ultrasensitive micromechanical
cantilevers, nanomagnetic tips, and displacement transducers -- could
also find new applications outside the area of spin detection.

\section{Conclusion}
\label{Conclusion}

Over the last two decades, MRFM has led to exciting progress in the
field of ultrasensitive spin detection and high-resolution MRI
microscopy.  Starting with early demonstrations in the 1990s imaging
with resolutions of a few micrometers -- on par with conventional MRI
microscopy -- the technique has progressed to the point where it can
resolve single virus particles and molecular monolayers.  Given the
fast pace at which modern nanofabrication technology is evolving, an
improvement of the method down to one-nanometer resolution seems
feasible without major changes to the instrument.  This resolution,
which is comparable to what three-dimensional electron microscopy
reaches on biological specimens, would be sufficient to map out the
coarse structure of many macromolecular complexes.  The extension of
MRFM to atomic resolution, where atoms in molecules could be directly
mapped out and located in 3D, remains an exciting if technically very
challenging prospect.

\begin{acknowledgments}
  The work discussed in this review was only possible because of the
  many fruitful experimental collaborations with H. J. Mamin and D.
  Rugar of the IBM Almaden Research Center.  The authors also thank
  these colleagues for their many detailed comments and very helpful
  discussions pertaining to this review.
\end{acknowledgments}

\end{document}